# Machine Learning enables Design of On-chip Integrated Silicon T-junctions with footprint of 1.2 μm × 1.2 μm

Sourangsu Banerji[1], Apratim Majumder[1], Alexander Hamrick[2], Rajesh Menon[1], and Berardi Sensale-Rodriguez[1, *]

[1] Department of Electrical and Computer Engineering, The University of Utah, Salt Lake City, UT 84112, USA

[2] School of Computing, University of Utah, Salt Lake City, UT 84112, USA

* E-mail: berardi.sensale@utah.edu

**Abstract**

To date, various optimization algorithms have been employed to design and improve the performance of nanophotonic structures. Here, we propose to utilize a machine-learning algorithm viz. binary-Additive Reinforcement Learning Algorithm (b-ARLA) coupled with finite-difference time-domain (FDTD) simulations to design ultra-compact and efficient on-chip integrated nanophotonic 50:50 beam splitters (T-junctions). Here we present the design of two T-junction splitters each with a footprint of only 1.2 μm × 1.2 μm. To the best of our knowledge, these designs are amongst the smallest ever reported till date across either simulations or experiments. The simulated net power transmission efficiency for the first T-junction design is ~ 82% and the second design is ~ 80% at λ = 1.55 μm. We envision that the design methodology, as reported herein, would be useful in general for designing any efficient integrated-photonic device for optical communications systems.

## 1. Introduction

In the last decade, the field of silicon nanophotonics has witnessed major breakthroughs [1]. The critical enabler for its unprecedented success can be attributed to the development of advanced foundry services [1, 2]. In addition to this, nanophotonic designs have been demonstrated with very efficient structures that can be implemented to obtain favourable characteristics like high sensitivity, low-loss, and high index contrast in dielectric distribution [3, 4, 5]. The integration of such all-dielectric passive nanophotonic components such as multiplexing couplers, waveguides, and so on with active devices such as lasers, LEDs, etc. onto a single chip will ultimately lead to the miniaturization of optical circuits with high data processing capability, very similar to what we see in silicon chips used for integrated electronics as of today. However, contrary to electronic circuits, there is still a lack of effective design methodologies in nanophotonics [6, 7].

Traditional nanophotonic design strategies are based upon theoretical and scientific intuitions [8-11]. However, most of the time, it does not provide analytical solutions for complex nanophotonic structures and light manipulation behaviour [12]. In addition to this, device designs based on analytical methods may also not satisfy performance requirements like compactness, efficiency, bandwidth, and power transmission. For this reason, a wide variety of numerical approaches such as evolutionary algorithm [13], objective-first inverse-design algorithm [14-17], topology optimization [18], nonlinear-search algorithm [19-23], and direct-binary-search algorithm [24-25] have been implemented to design integrated-nanophotonic structures. Amongst all, inverse optimization, or objective-first inverse-design algorithms have been shown to deliver the best performing nanophotonic structures with adequate computational trade-offs [14–17, 26-32]. From such a perspective, we can see that inverse-design algorithms (one such example is the adjoint method) are highly suitable for developing next-generation, compact nanophotonic devices with novel functionalities and features.

Machine learning has also recently emerged and attracted a great deal of attention from both academia and industry alike as a viable design methodology. In all areas of physics itself, ranging from gravitational wave analysis [33], to materials designs [34, 35], to phase transitions in quantum physics [36, 37], machine learning have successfully been leveraged to provide for performance comparable to some of the most advanced design methodologies in a natural and straightforward manner. In all these previous examples, we observe that the advantage of machine learning lies in the accurate modelling and characterization of complex relationships within the underlying systems. To summarize, the advantages of machine learning are four-fold. First, machine learning algorithms allow for *hardware parallelization*. For example, if we consider popular evolutionary algorithms, we will observe that they heavily depend on two important factors: (a) the number of generations, and (b) diversity of initial solutions. In fact, their computational complexity rises with each necessary operation like reproduction, mutation, recombination, selection, and survival of the fittest [38]. Combination of these operations adversely affects the performance, probabilistic transition, and convergence for these algorithms.

In contrast, machine-learning algorithms, even the one described in this paper, do not require any operation of such sort. The algorithm described in this paper is fully parallelizable. The training phase can be divided, and the simulations can be distributed arbitrarily across multiple computers, i.e., the data is generated and evaluated independently of each other [12, 38]. *This is true not just in case of genetic algorithms only but for all class of optimization algorithms, including inverse*

*optimization or even topology optimization. By transitioning from an optimization-based design methodology to a prediction-based one, we gain computational advantage via hardware parallelization* [39].

Second, machine learning also does not depend on the quality of initial solutions to guarantee reasonable solutions. Third, in contrast to an inverse-design algorithm like the adjoint method, machine learning can solve the forward design problem much faster with a neural network (deep learning). Even though this advantage is not big enough when comparing against the adjoint method, which requires only two forward simulations for the entire optimization, machine learning still has a marginal advantage in the sense that one can restrict the design space to manufacturable devices and physical solutions, which are harder to find with adjoint methods. Fourth, in machine learning (especially in deep learning), the model is trained to "***intelligently learn***" the non-linear relationships between the input and output over a large dataset. The model in this way can also "***intelligently learn***," for example, Maxwell's equations and solve them, without explicitly knowing about them. This allows for possible discovery of solutions outside of the boundaries of the training data, and the ability to transfer knowledge between problems by a method known as "*transfer learning*." This approach represents a complete paradigm shift in thinking of how nanophotonics research has been understood till date and what it could lead to in the time to come; to enable equally disruptive series of novel findings in nanophotonics.

Considering the advantages stated above, researchers working in the field of optics and photonics have started harnessing machine learning to develop foundry compatible optical components for large scale industrial rollout [39-45]. In this work, we utilized a machine learning algorithm, namely binary-Additive Reinforcement Learning Algorithm (b-ARLA) coupled with a finite-difference time-domain (FDTD) method to demonstrate efficient and ultra-compact 50:50 beam splitters (T-junctions) as shown in Fig. 1(a). Out of various machine learning algorithms, there are many advantages of using reinforcement learning algorithms. First, reinforcement learning algorithms are generally used to solve very complex problems that cannot be solved by conventional techniques therefore it can achieve long-term results and can outperform humans in many tasks. In fact, DeepMind's AlphaGo, a notable reinforcement learning model, was shown to beat the current world champion at the game of Go in March 2016 [46]. Second, it can correct for the errors that occurred during the training process; therefore, once an error is corrected by the model, the chances of occurring the same error are very less [47, 48]. Third, in the absence of a training dataset, it is bound to learn from its experience, hence, it is intended to achieve the ideal behaviour (similar to a real human being) within a specific context, to maximize its performance [49]. Finally, reinforcement learning algorithms strikes a balance between exploration and exploitation. Exploration is the process of trying different things to check if they are better than what has been tried before whereas, exploitation is the process of trying the things that have worked best in the past. Other learning algorithms do not perform this balance [47, 48]. The top view of the "unit cell," as shown in Fig. 1(b) consists of square sub-unit pixels of either silicon or air. Square pixels have an advantage from the fabrication point-of-view. Although electron-beam-lithography can be used to fabricate the structures presented in this paper (minimum feature size ~ 100nm), for large scale fabrication of a large number of devices on wafers, typically the industry standard is to rely on immersion projection lithography. It is well-known that optical proximity correction is required to ensure correct size and shape of the features [50]. Since the optical lithography industry has for a long time perfected this process for square and rectangular features which are the most abundant geometrical shapes present in ICs, we believe that our structures can be readily adapted to multi-wafer processes. Hence, square pixels have an advantage during fabrication. In addition to this, circular pixels, i.e. holes may also be used to design similar digital metamaterial devices, as have been demonstrated elsewhere [51].

Finally, Table 1 provides a more detailed analysis of power splitters reports till date and compares it with the work as described in this paper. A few noteworthy implementations of ultra-compact Y- or T- junction 50:50 power splitters reported in the scientific literature have had area footprints > 2 μm × 2 μm with < 1dB insertion loss at the telecom wavelength of 1.55 μm [45, 52-54]. Therefore, it is evident that with a footprint of only 1.2 μm × 1.2 μm, the designs reported in this paper are amongst the smallest ever reported across either simulations or experiments as of date.

**Table 1: Summary of symmetric and asymmetric nanophotonic power splitters**

| Reference | Splitter shape | Imbalance (split ratio) | Insertion loss | Bandwidth (center wavelength = 1550 nm) | Polarization | Footprint |
|---|---|---|---|---|---|---|
| Tahersima et. al.[45] | Y | 50:50 | - | 200 nm | TE | 2.6 × 2.6 μm$^2$ |
| Tahersima et. al.[45] | Y | 25:75 | - | 200 nm | TE | 2.6 × 2.6 μm$^2$ |
| Zhang et. al. [52] | Y | 50:50 | 0.13 dB | ~ 80 nm | TE | 1.2 × 2 μm$^2$ |
| Kurt et. al.[53] | T | 50:50 | < 0.3 dB | ~200 nm | TM | 4.096 × 4.096 μm$^2$ |
| Alpkilic et. al. [54] | T | 100:0 | < 0.3 dB | - | TM | 2.8 × 2.8 μm$^2$ |
| Alpkilic et. al. [54] | T | 100:0 | 0.73 dB | - | TE | 2.8 × 2.8 μm$^2$ |
| Xu et. al.[55] | Y | 50:50 | ~ 1dB | ~30 nm | - | 3.6 × 3.6 μm$^2$ |
| Xu et. al.[55] | Y | 40:60 | ~ 1dB | ~30 nm | - | 3.6 × 3.6 μm² |
| Xu et. al.[55] | Y | 25:75 | ~ 1dB | ~30 nm | - | 3.6 × 3.6 μm² |
| Lin et. al.[56] | Y | 50:50 | 0.36 dB | ~100 nm | TE | 1.4 × 2.3 μm² |
| Ren et. al.[57] | Y | 50:50 | 0.33 dB | ~ 40 nm | TE | 2.4 × 3 μm² |
| Xie et. al.[58] | Y | 50:50 | < 1.5 dB | ~ 40 nm | TE | 4 × 4.5 μm² |
| Chang et. al.[59] | Y | 50:50 | < 1.5 dB | ~ 80 nm | TE | 2.88 × 2.88 μm² |
| This work | T | 50:50 | 0.86 dB | ~ 40 nm | TE | 1.2 × 1.2 μm² |
| This work | T | 50:50 | 0.95 dB | ~ 40 nm | TE | 1.2 × 1.2 μm² |

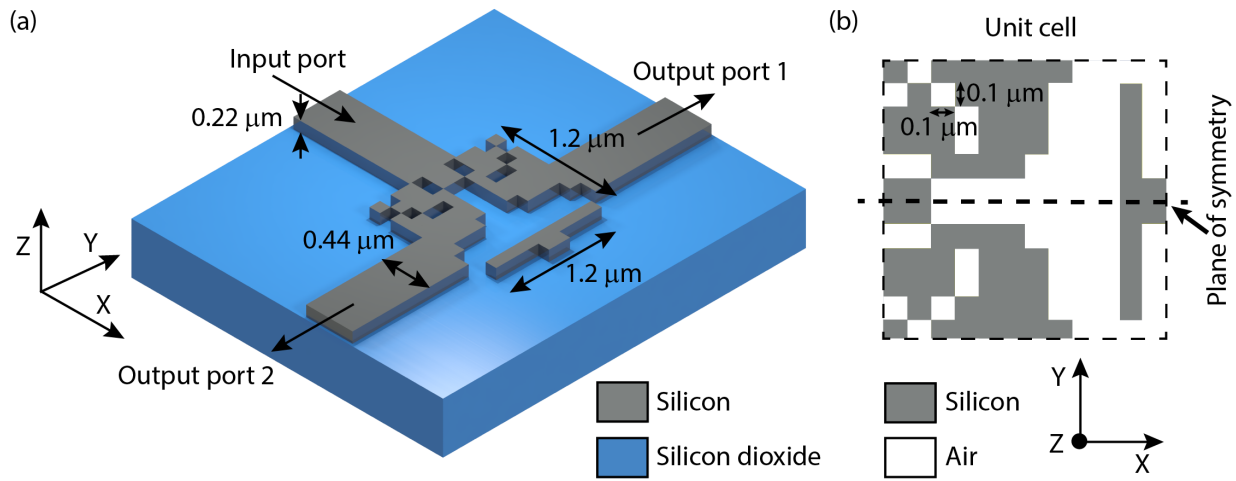

Fig. 1. Schematic of a pixelated nanophotonic structure. (a) 3D representation of the T-junction splitter. (b) Top view of the unit cell.

## 2. Design and Optimization

The perceptron-like machine learning algorithm [55] used in this study is implemented to reduce the insertion loss (splitting the input power with minimal loss) of the power splitter at an operating wavelength of 1.55 μm. The algorithm developed herein combines both the "*additive updates*" feature of a perceptron algorithm [54, 60] as well as the "*reward for state idea*" of reinforcement learning [60]. The flowchart of the algorithm is shown in Fig. 2, which depicts that the algorithm consists of two phases: training and inference.

## 2.1. Training Phase

The training phase starts with creating a photonic structure where each constituent sub-unit of the "unit cell" is randomly distributed. Essentially, the entire "unit cell" consists of 12 x 12 randomly distribution pixels. However, due to the inherent symmetry of the structure itself, i.e., 50:50 split ratio, the generated random structure now consists of 12 x 6 binary cells (flip symmetry across the y-direction), where "1" denotes the high refractive index Si- sub-units and low refractive index air- sub-units are represented by "0" in the binary "unit cell." Keeping in mind the capability of current fabrication technologies, the design parameters of each sub-unit within the "unit cell" are taken very conservatively (e.g., 100 nm minimum features). Therefore, the initial design parameters are fixed as follows: square-shaped sub-units with a size of 0.1 μm × 0.1 μm, the height of the structure is 0.22 μm (typical in SOI), and the material refractive index is $n_{Si}$ = 3.46. The refractive index of air is $n_{air}$ = 1. Therefore, the size of the complete photonic structure is 1.2 μm × 1.2 μm × 0.22 μm. The size of the input and output waveguides is fixed at 1 μm (length) × 0.44 μm × 0.22 μm (cross-section). The substrate thickness is assumed to be 0.6 μm. Following the creation of the "unit cell" structure along with the input and the output waveguide ports, a 2.5D varFDTD (variational FDTD) method [61] is incorporated to analyse the time-domain response of the photonic structure at λ = 1.55 μm. Common desktop CPUs were employed to perform the 2.5D varFDTD simulations (Lumerical Inc.'s Lumerical MODE solutions). A more elaborate description of the full-wave simulation is provided in the following section. While extracting the response of the structure, the insertion loss (in dB) was extracted for the nanophotonic structure.

The rationale behind considering the insertion loss as the metric for optimization is inspired by the fact that photonic integrated circuit designers would typically be interested in the total amount of light that is transmitted into the fundamental mode. Hence, the formal definition of "Insertion loss" is the loss of signal power from the insertion of the T-junction from the input port to each output port. Mathematically, this can be as:

$$\text{Insertion Loss (I. L.)} = -10\log_{10}(\text{mean}(T_{net})) \tag{1}$$

where $T_{net}$ is the net transmission into the fundamental mode of the output waveguides. The "-ve" sign in eqn. (1) is omitted while reporting the insertion loss values in the results and discussion section later to keep it uniform with how the values are reported in the literature for easy comparison [45, 52-59]. The difference between the numerical value of the insertion loss in the worst possible scenario and the insertion loss extracted from the steady state response of the $i^{th}$ randomly generated photonic layout is defined as the reward function $R_i$, which is then defined as follows:

$$R_i = |I.L._{worst} - I.L._i| \tag{2}$$

In the above expression, $I.L._{worst}$ as stated earlier of the insertion loss in the worst possible scenario, which essentially refers to two different scenarios: a binary "unit cell" all of whose individual sub-unit elements are either (a) "0" (air) or (b) "1" (silicon). The reward function $R_i$ approaches to $I.L._{worst}$ only when $I.L._i \sim 0$, i.e., when any of the $i^{th}$ iteration results in a structure where the splitting of the input power is ideally lossless.

The modulus of the reward function $R_i$ generates a single valued positive number which is then multiplied with the $i^{th}$ binary square matrix ($B_i$). One can think of this step as assigning a "*confidence*" score to the all the pixels within $B_i$. Mathematically, this can be expressed as,

$$R_i \times \begin{bmatrix} a_{11} & \cdots & a_{j1} \\ \vdots & \ddots & \vdots \\ a_{1k} & \cdots & a_{jk} \end{bmatrix} \quad (3)$$

where, $B_i = \begin{bmatrix} a_{11} & \cdots & a_{j1} \\ \vdots & \ddots & \vdots \\ a_{1k} & \cdots & a_{jk} \end{bmatrix}$, such that $(a_{11},..,a_{j1},..,a_{1k},..,a_{jk}) \in (0,1)$. For the sake of this problem, we consider $j = k$. This will simplify to a single matrix, say reward matrix $RM_i$ since $R_i$ is a scalar as shown below,

$$RM_i = \begin{bmatrix} R_i a_{11} & \cdots & R_i a_{j1} \\ \vdots & \ddots & \vdots \\ R_i a_{1k} & \cdots & R_i a_{jk} \end{bmatrix} \quad (4)$$

or, as $RM_i = R_i B_i$. This is done for all the "n" number of randomly generated binary square matrices as shown within the training phase block of Fig. 2. Next, all the "n" number of such matrices are summed together into a single matrix which is known as the cumulative reward matrix (CRM). The CRM can be now written as,

$$CRM = RM_1 + RM_2 + \cdots + RM_n = \sum_i^n RM_i \quad (5)$$

The CRM is of the same dimension as the reward matrix $RM$ or the binary square matrix $B$. It contains the accumulated rewards for each pixel i.e. the algorithm now has the information of the specific positions within the square binary matrix which has one of the highest "*confidence*" scores. This will be beneficial later. The CRM is then saved and passed onto the final phase: The inference phase.

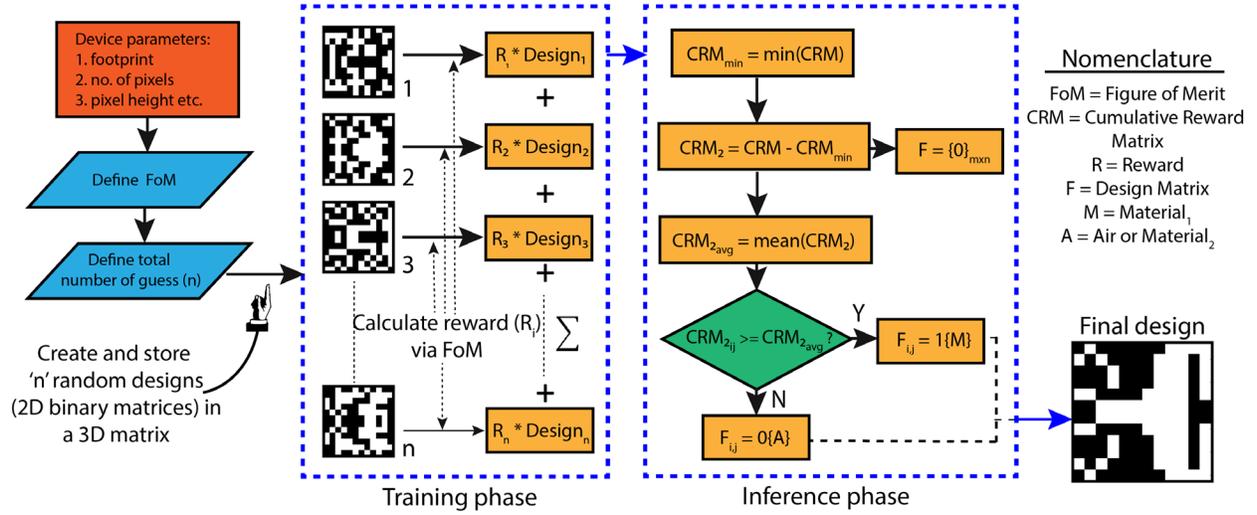

Fig. 2. Flow diagram of the binary-Additive Reinforcement Learning Algorithm (b-ARLA)

2.2. Inference Phase

The CRM helps the algorithm to generalize the problem. A simple way to understand this is as follows. There are 2 matrices $B_i$ and $B_{i+1}$ with reward value of $R_i$ and $R_{i+1}$. Here, for simplicity $R_i \geq R_{i+1}$. The $j^{th}$ position of both $B_i$ and $B_{i+1}$ can contain (1) both "1", (2) both "0" and (3) combination of "1" and "0". The accumulated reward for the $j^{th}$ position in the CRM will be (a) $(1 \times R_i + 1 \times R_{i+1})$, (b) $(0 \times R_i + 0 \times R_{i+1})$, and (c) $(0 \times R_i + 1 \times R_{i+1})$ or $(1 \times R_i + 0 \times R_{i+1})$. Simple analysis tells that (a) > (c) > (b). The algorithm understands the positions of "1" and "0" based on this "*confidence*" scores in the CRM. Finally, what remains is too just to filter and separate it out. Therefore, the first step is to ascertain this minimum. This is then subtracted from the CRM. In, parallel, a final binary matrix F is created which is initialized with all "0"'s. This takes care of "0" valued pixels in the matrix. In the subsequent step, the mean of the CRM is taken to handle for cases like case (1) and case (3) as shown in the former example because a value greater than this mean only guarantees that specific pixel to be "1" without fail as the cumulative reward values for such pixel positions will be amongst the highest. Nonetheless, the rationale behind evaluating the mean here can also be understand from the perspective of signal processing where signal averaging is carried out to extract the real signal from a noisy channel. Hence, the final decision-making step involves this check. If the value is found to be greater than the mean, the matrix F is updated with a value of "1" else is kept as it is i.e. "0".

## 3. Results and Discussion

The designed nanophotonic structure comprises of square pixelated Si- sub-units that are intelligently distributed in an air medium, i.e., the algorithm predicts the location of the Si- sub-units to obtain the desired 50:50 power splitting of the input power with minimal loss. The T-junctions are designed for a light source with the fundamental TE polarization (with non-zero components of $E_x$, $E_y$ $and$ $H_z$). To avoid any undesired back reflections, perfectly matched layers (PML) surrounded the boundaries of the computational domain.

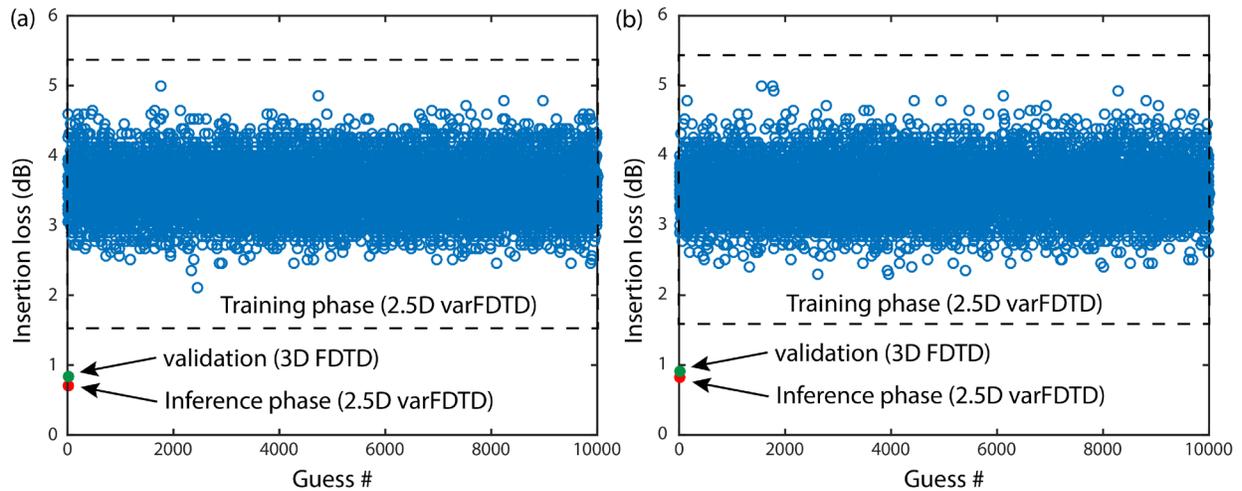

Fig. 3. The insertion loss (in dB) for each guess (hollow blue dots) and final prediction (solid red dot) using 10000 guesses across both designs. The minimum possible insertion loss achieved in the final prediction from the inference phase across both designs are (a) ~ 0.86 dB and (b) ~ 0.95

dB respectively, although the insertion loss (in dB) for nearly all guesses during the training phase for both (a) and (b) are in the range ~ 2 to 5 dB.

Two different commercially available software from Lumerical Inc. were employed during the entire process. During the training as well as the inference phase, Lumerical MODE solutions was utilized to extract the response (value of insertion loss) of the nanophotonics structure via 2.5D varFDTD. An additional post-validation after the inference phase was employed to cross-check the obtained results across a full 3D FDTD with Lumerical FDTD solutions. The reason behind using 2.5D varFDTD was to speed up the "*learning*" process. The 2.5D varFDTD, as well as the 3D FDTD, had a mesh accuracy of 1/35 of the free space wavelength to ensure accurate modelling. Parallelization during the training phase for the 2.5D varFDTD simulations was carried out on 10 Intel Pentium i7 CPUs with 16 GB RAM each. The complete numerical simulations took ~ 67 hours (almost three days). The number of matrices used in the learning phase (i.e., N rounds) was empirically set to 10000. Care was also taken to make sure that this randomly generated binary "unit cell" matrix is unique across all the CPUs such that the same structure is not running on two different CPUs at any given instant of time.

Two different designs for the on-chip integrated nanophotonic in-plane incidence 50:50 T-junctions were made, each corresponding to a different worst possible insertion loss (I.L.$_{worst}$) scenario as derived in Eqn. (2). The 2.5D FDTD simulated I.L.$_{worst}$ value for a binary "unit cell," all of whose individual sub-unit elements are "0" (air) was ~ 9.77 dB and a binary "unit cell" all of whose individual unit elements are "1" (silicon) was ~ 16.62 dB. Theoretically speaking, even though this change in I.L.$_{worst}$ value would neither have changed anything in relation to the working principle of the algorithm nor the final prediction, it was certainly interesting to cross-check this before making any conclusive statements. As observed in Fig. 3, the insertion loss for each guess and final prediction using 10000 guesses follows a very similar trend. For nearly all guesses during the training phase, the average insertion loss was in the range of ~ 2 to 5 dB. The minimum insertion loss was achieved in the final prediction from the inference phase across both the designs. The insertion loss (efficiency in terms of total power-out as % of power-in) was determined to be ~ 0.82 dB and ~ 0.87 dB across 2.5D varFDTD for both the designs, respectively. The full 3D FDTD gives a slightly worse but more accurate value for the insertion loss at ~ 0.86 dB and ~ 0.95 dB for the operational wavelength of 1.55 μm for each design, respectively. The subsequent "unit cell" structure, steady-state electric field distribution at $\lambda = 1.55$ μm, and the insertion loss for each design under broadband operation (1.45 - 1.65 μm) are plotted in Fig. 4(a-b). It can be observed that for both the designs that the insertion loss was to some extent, virtually wavelength insensitive with variations below 10% over the wavelength range from 1.52 μm to 1.57 μm. To be specific, the bandwidth was from ~ 1.53 μm to ~ 1.57 μm (~40nm bandwidth) for the first design in Fig. 4(a). For the other design in Fig. 4(b), this bandwidth was in the range of 1.52 μm to 1.56 μm (~ 40nm bandwidth). One must keep in mind here that the machine learning algorithm was trained on and the final inference was made at only a single wavelength of 1.55 μm. There is no crosstalk amongst the output waveguides. Scattering was negligible, as evidenced by the steady-state response plots in Fig. 4(a-b) with the appearance of an interference pattern at the input end, indicating the existence of very weak back-reflection. Furthermore, upon even more close inspection of the steady-state intensity profile in Fig. 4(a-b) one can further conclude : (a) an efficient splitting of the fundamental mode and (b) a strong modal match at the output port (waveguide) with excellent coupling efficiency for both the splitter designs.

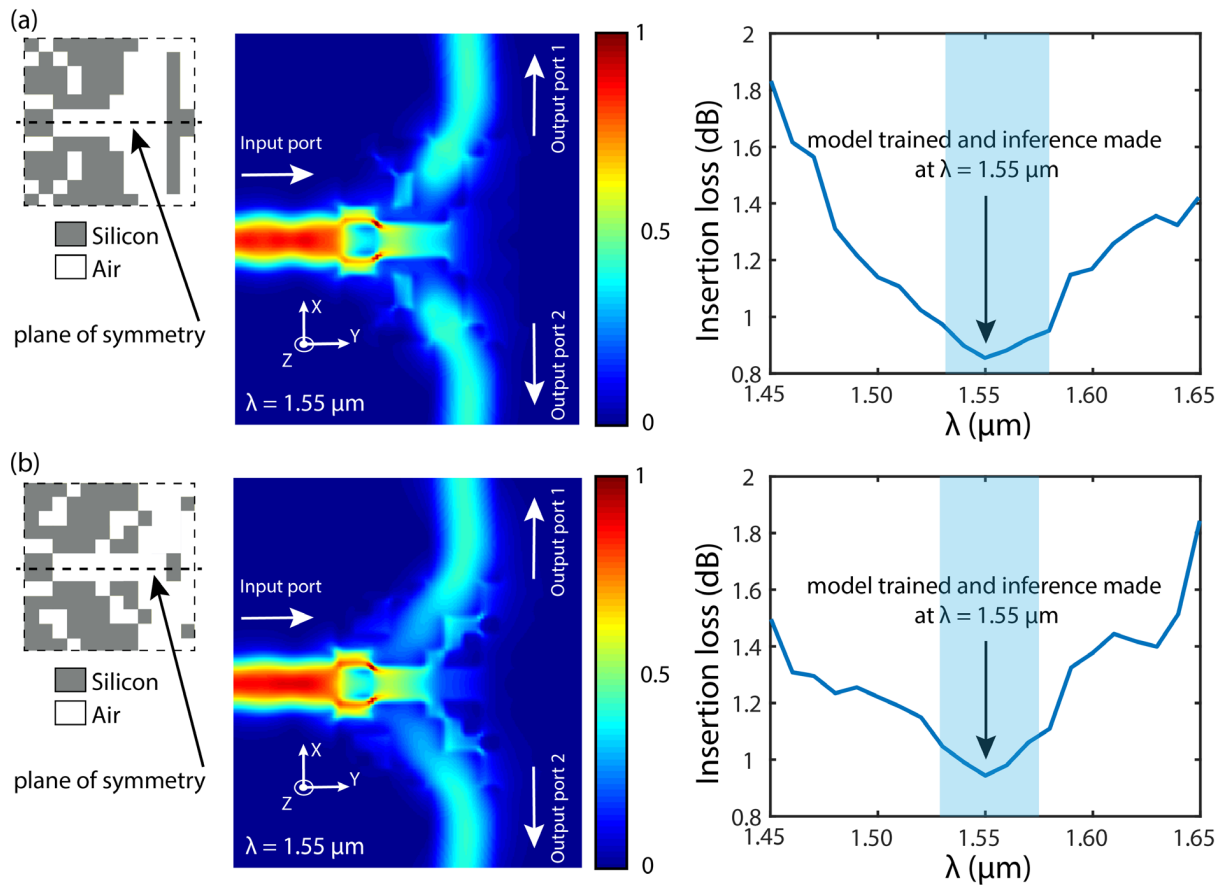

Fig. 4. The predicted pixel profile, steady-state response at the operational wavelength of 1.55 μm and the insertion loss of the structure under broadband operation (1.50 - 1.60 μm) for the design where I.L.worst refers to a binary "unit cell" all of whose individual sub-unit elements are either (a) "0" (air) or (b) "1" (silicon).

Finally, an unbiased comparison in relation to large conventional integrated beam splitters needs to be carried out to highlight the true significance of this work, apart from the design methodology discussed in the previous sections. We acknowledge the fact here that the designs discussed in this study are certainly not the best in terms of power efficiency (insertion loss) compared to what has already been reported in the literature. An insertion loss of 0.8 to 1 dB (as in the case of the designs reported herein) in general corresponds to a net power transmission efficiency of ~ 80-83%. Conventional integrated beam splitters or even optimized 50:50 splitters (Y-shape or T-shape) report insertion losses of < 0.3 dB (~90-93% in terms of net power transmission) [2, 38, 45, 52-54]. However, we believe that the advantage of a small area footprint outweighs such a marginal ~ 10% reduction in efficiency from the perspective that now a greater number of devices can be integrated together in a single photonic chip than what has been previously possible (analogous to what has been seen for transistors in electronic circuits over the last decades). This will eventually lead designers to design complicated photonic logic circuits with more flexibility. We can proclaim this as a "*Photonic Moore's law*" [62]. This trade-off between efficiency and footprint can be handled by utilizing "unit cell" structures with a larger number of sub-unit pixels in the same total area, that is a larger sub-unit density (say for example 30 x 30 or even 60 x 60) to approximate sharp bends more gradually than what has been done with a coarse structure (12 x 12) in this paper. The geometry constraints assumed in this work were taken so to represent what can be demonstrated at a standard university-level fabrication facility without much difficulty. However, at industrial foundries, one can expect to exploit this to create even more efficient beam splitter structures with such an ultra-compact area footprint. Apart from this, the use of higher refractive index materials [21, 23] has also been previously shown to improve device performance and is expected to provide the same advantages here as well.

In addition to this, a smaller structure will also have a lower heat generation "*per device*" in contrast to a conventional one [63]. Now, if one is interested in lowering the overall heat generation in the photonic circuit keeping the total number of individual structures same, a smaller footprint will have the advantage that heat sinks can be easily accommodated within the same area along with the device in place of a conventional larger structure taking up the same amount of space. This will eventually lower the *operational cost* and provide cost-effective solutions [64, 65].

## 4. Conclusion

We introduced the design of subwavelength ultra-compact and efficient on-chip integrated nanophotonic 50:50 beam splitters (T-junctions) via a machine learning algorithm. We numerically investigated its power splitting effect at an operating wavelength of $\lambda = 1.55$ μm by using FDTD simulations. Despite its low insertion loss, as indicated from the full-wave simulations, we would like to point out that imperfections and impurities during the fabrication step will inevitably decrease the device efficiency and degrade its performance. However, these problems will not have any detrimental effect in relation to the functionality of the designed structure. From such perspective, using the algorithm as described in this paper one can now extend this to realize situation dependent nanophotonic splitter designs such as asymmetric power splitters (which are optimized for unbalanced power splitting i.e. 40:60, 20:80 or 100:0), splitters based on different geometries, for example Y-branch splitters or even polarization beam splitters by defining an appropriate reward function. Nonetheless, reconfigurability can also be achieved with the use of active materials in these pixelated "logic cells". Overall, the method employed and our results evidence that the use of machine learning algorithms is a promising technique for the inverse design of wide variety of efficient passive and active integrated photonics devices.

**Acknowledgments**

This work was supported by the NSF awards: ECCS # 1936729 and MRI #1828480